\documentclass[aps,prl,showpacs,twocolumn,floatfix,nofootinbib,superscriptaddress]{revtex4}
\usepackage{graphicx} 
\usepackage{amsmath}
\usepackage{bm}
\usepackage{longtable}
\usepackage{dcolumn}
\begin{document}

\title{
Mapping out atom-wall interaction with atomic clocks
}

\author{A. Derevianko}
\affiliation{Department of Physics, University of Nevada, Reno,
Nevada 89557, USA}

\author{B. Obreshkov}
\affiliation{Department of Physics, University of Nevada, Reno,
Nevada 89557, USA}

\author{V. A. Dzuba}
\affiliation{Department of Physics, University of Nevada, Reno,
Nevada 89557, USA}

\affiliation {
School of Physics, University of New South Wales, Sydney,
2052, Australia}


\date{\today}
\begin{abstract}
We explore a feasibility of measuring atom-wall interaction using atomic clocks based on atoms trapped in engineered optical lattices.  Optical lattice is normal to the wall.
By monitoring the wall-induced clock shift at individual wells of the lattice,
one would measure a dependence of the atom-wall interaction on the atom-wall separation.
We rigorously evaluate the relevant clock shifts and show that the proposed scheme may uniquely probe the long-range
atom-wall interaction in all three qualitatively-distinct regimes of the interaction: van der Waals (image-charge interaction), Casimir-Polder (QED vacuum fluctuations)  and Lifshitz (thermal bath fluctuations). The analysis is carried out for atoms Mg, Ca, Sr, Cd, Zn, and Hg, with a particular
emphasis on Sr clock.
\end{abstract}


\maketitle

Atomic clocks define the unit of time, the second. Usually the environmental effects (e.g., stray fields) degrade the performance of the clocks. One may turn this around and by measuring shifts of the clock frequency, characterize an interaction with the environment. The most fundamental experiments of this kind search for a potential variation of fundamental constants~\cite{ForAshBer07}, where the ``environmental agent'' is the fabric of the Universe itself, affecting the rate of ticking of atomic clocks. In this paper, we evaluate a feasibility of using atomic clocks to measure basic laws of atom-wall interactions. We find that a certain class of atomic clocks, the optical lattice clocks, are capable of accurately characterizing the atom-wall interaction.
Moreover, this is a unique system where the atom-wall interaction may be probed in all three qualitatively-distinct regimes of the interaction: van der Waals (image-charge interaction), Casimir-Polder (QED vacuum fluctuations)  and Lifshitz (thermal bath fluctuations).

Understanding the basic  atom-wall interaction~\cite{BloDuc05} is important, for example, for probing
a hypothetical ``non-Newtonian'' gravity at a $\mu$m scale (see e.g., Ref.~\cite{Ran02}).
Also, with miniaturization of atomic clocks, for example, using atomic chips~\cite{GalHofSch09},
the  atom-wall interaction will become an important systematic issue.

In optical lattice clocks, ultracold atoms are trapped in minima (or maxima) of intensity of a
standing wave of a laser light (optical lattices) operated at a
certain ``magic'' wavelength~\cite{KatTakPal03,YeKimKat08}. The
laser wavelength is tuned so that the differential light
perturbations of the two clock levels vanishes exactly.
Such ideas
were experimentally
realized~\cite{TakHonHig05,LeTBaiFou06,LudZelCam08etal} for optical
frequency clock transitions in divalent atoms, such as Sr, yielding fractional accuracies
at a $10^{-16}$ level~\cite{LudZelCam08etal}.  The clock transition is between the ground $^1S_0$ and the lowest-energy
excited $^3P_0$ state. $J=0$ spherical symmetry of the clock states makes the clock insensitive to stray magnetic fields and
environmentally-induced decoherences.

\begin{figure}[h]
\begin{center}
\includegraphics*[scale=0.25]{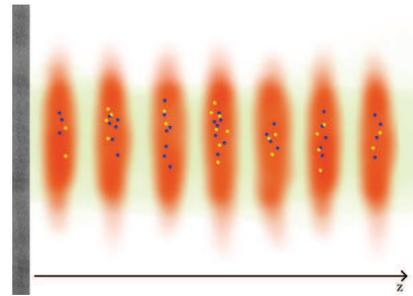}
\caption{(Color online) Idealized setup for measuring atom-wall interaction with optical lattice clocks.
Clouds of ultracold atoms are trapped in an optical lattice operating at  a ``magic'' wavelength.
By monitoring the wall-induced clock shift at individual trapping sites,
one measures a dependence of the atom-wall interaction on the atom-wall separation.
\label{Fig:setup}}
\end{center}
\end{figure}

An idealized setup for measuring atom-wall interaction is shown in Fig.~\ref{Fig:setup}.  The conducting surface of interest acts as a mirror for the laser beam normally incident on the surface. The resulting interference of the beams forms an optical lattice. Laser operates at a ``magic'' wavelength $\lambda_m$ specific to the atom  (see Table~\ref{Tab:atoms}). For all tabulated magic wavelengths, atoms are attracted to maxima of the laser intensity and one could work with 1D optical lattices. The first pancake-shaped atomic cloud would form at $\lambda_m/4$ distance from the mirror. The subsequent adjacent clouds are separated by a distance $\lambda_m/2$.  Ref.~\cite{SorAlbFer09} discusses an experimental procedure for loading atoms into sites close to a mirror.

 An  advantage of working with optical lattices lies with a tight spatial confinement at the lattice sites. Commonly, the accuracy of determination of the atom-wall interaction is limited by the spatial extent of an atomic ensemble~\cite{BloDuc05}. In optical lattices, the size of ultracold atom wave-function can be reduced to a small fraction of the lattice laser wavelength. 

\begin{table*}[h]
\caption{Fractional shifts of the ${^1}S_0-{^3}P_0$ clock transitions
in divalent atoms due to presence of an ideal conducting surface.
The second column lists clock frequencies. Values of magic wavelengths, $\lambda_m$, in Sr and Yb
are experimental~\cite{TakKat03,BarHoyOat06}  and are our theoretical results for other atoms.
The differences of the static polarizabilities $\alpha(0)$ and the van der Waals coefficients $C_3$ for the two clock levels are tabulated in the fourth and fifth columns.
Finally, we list the fractional-shift parameters $\beta_{CP}$, $\beta_{vdW}$ and $\beta_L$, Eq.~(\ref{Eq:beta}).  Notation $a[b]$  stands for $a \times 10^b$. \label{Tab:atoms}}
\begin{ruledtabular}
\begin{tabular}{cccccccc}
atom  &  $\nu_{clock}$, Hz &     $\lambda_m$, nm   & $\Delta \alpha(0)$, a.u. &  $\Delta C_3$, a.u. &    $\beta_{vdW}$ &   $\beta_{CP}$ &         $\beta_L$  \\
\hline

Mg &   6.55[14] & 466 &  29 & 0.21 &  -3.1[-12] &   -7.9[-13] &   -1.0[-13]  \\
Ca &   4.54[14] & 739 & 138 & 0.17 &  -8.8[-13] &   -8.6[-13] &   -1.8[-13]  \\
Sr &   4.29[14] & 813 & 261 & 0.25 &  -1.1[-12] &   -1.2[-12] &   -2.6[-13]  \\
Yb &   5.18[14] & 759 & 155 & 0.35 &  -1.5[-12] &   -7.6[-13] &   -1.6[-13]  \\
Zn &   9.69[14] & 416 &  28 & 0.30 &  -4.2[-12] &   -8.2[-13] &   -9.4[-14]  \\
Cd &   9.03[14] & 419 &  28 & 0.31 &  -4.6[-12] &   -8.7[-13] &   -1.0[-13]  \\
Hg &   1.13[15] & 362 &  22 & 0.30 &  -5.5[-12] &   -9.8[-13] &   -9.8[-14]  \\
\end{tabular}
\end{ruledtabular}
\end{table*}

Two earlier proposals, by Florence~\cite{SorAlbFer09}  and Paris~\cite{WolLemLam07} groups,
considered trapping divalent atoms in optical lattices for studying atom-wall interaction.
In both proposals the lattices are oriented vertically and ultracold atoms experience a combination of periodic
optical potential and linear gravitational potential. In the Florence proposal~\cite{SorAlbFer09}, the atom-wall
interaction modifies Bloch oscillation frequencies of atomic wavepackets in this potential.
In the Paris proposal\cite{WolLemLam07},  laser pulses at different frequencies
are used to create an interferometer with a coherent superposition of atomic wavepackets at different sites.
Here we consider an alternative: by monitoring the clock shift at individual trapping sites,
one measures a distance dependence of the atom-wall interaction.

{\em Qualitative estimates -- }
 As the separation $z$ between an atom and a wall increases, the atom-wall
interaction evolves  through several distinct regimes: (i) chemical-bond
region that extends a few nm from the surface, (ii) van der Walls region,
(iii) retardation (Casimir-Polder) region, and (iv) the thermal (Lifshitz) zone.
The chemical-bond region
is beyond the scope of our paper and we focus on the three longer-range regimes
of the interaction between a perfectly conducting wall and a
spherically-symmetric atom.

Qualitatively, the van der Waals interaction arises due to an interaction of
atomic electrons and nucleus with their  image charges
\begin{equation}
U_{vdW}\left(  z\right)  =-C_{3}\, z^{-3},\label{Eq:vdWlim}%
\end{equation}
where  the coefficient $C_{3}$ depends on an atomic state. It may be expressed in
terms of the electric-dipole dynamic polarizability of the atom as
\begin{equation}
C_{3}=\frac{1}{4\pi}\int_{0}^{\infty}\alpha\left(  i\omega\right)  d\omega \,.
\label{Eq:C3}
\end{equation}

Eq.(\ref{Eq:vdWlim}) assumes instantaneous exchange of virtual photons. More
rigorous consideration in the framework of  QED leads to
the Casimir-Polder limit~\cite{CasPol48}
\begin{equation}
U_{CP}\left(  z\right)  =-3/(8\pi) \, \hbar c~\alpha\left(  0\right)
~{z^{-4}}.\label{Eq:CPlim}%
\end{equation}
Notice the appearance of the speed of light $c$ in this formula. A
transition between the van der Walls and the retardation regions occurs at the
length-scale $\hbar c/\Delta E_{a}$, where $\Delta E_{a}$ is a characteristic
value of the atomic resonance excitation energy. Compared to the  van der
Waals interaction, the retardation potential  has a steeper, $z^{-4}$,
dependence on the  atom-wall separation.

The Casimir-Polder interaction, Eq.(\ref{Eq:CPlim}), is mediated by vacuum
fluctuations of electromagnetic field. At finite temperatures $T$,
populations of the vacuum modes are modified and a new length-scale, $\hbar
c/(k_{B}T)$, appears. As shown by Lifshitz~\cite{Lif56}, the distance dependence of the interaction
switches back  to the  inverse cubic dependence of the van der Waals
interaction, Eq.(\ref{Eq:vdWlim}),
\begin{equation}
U_{L}\left(  z\right)  =-{1}/{4} \, k_{b}T~\alpha\left(  0\right)  ~
{z^{-3}} \, .\label{Eq:Tlim}
\end{equation}

Due to the interaction with the wall, both clock levels would shift. We may parameterize the resulting fractional clock shifts as
%
\begin{equation}
\frac{\delta\nu}{\nu_\mathrm{clock}}\left( z, T\right) =\left\{
\begin{array}
[c]{c}%
\beta_{vdW}\, \left(  \frac{\lambda_{m}}{z}\right)  ^{3} \, , \\
\beta_{CP} \, \left(  \frac{\lambda_{m}}{z}\right)  ^{4} \, ,\\
\beta_{L} \, \left(  \frac{T}{300K}\right)  \left(  \frac{\lambda_{m}}{z}\right)^{3} \, .
\end{array}
\right.
\label{Eq:beta}
\end{equation}
We evaluated coefficients $\beta$ for the clock transitions in Mg, Ca,
Sr, Yb, Zn, Cd, and Hg (see discussion of the method later on). The results are presented in Table~\ref{Tab:atoms}.
The estimates of Table~\ref{Tab:atoms} immediately show that the atom-wall interaction is a large effect, corresponding to $10^{-10}$ fractional clock shifts at the first well. This is roughly a millon time larger than the demonstrated accuracy of the Sr clock~\cite{LudZelCam08etal}.


{\em Rigorous consideration --- }
In general, as the atom-wall separation is varied, there is a
smooth transition between the three interaction regimes. To properly describe the
crossover regions, we employ an expression by \citet{BabKliMos04}, which may be
represented as
\begin{equation}
U\left(  z,T\right)  =-\frac{k_{B}T}{4z^{3}}\left[  \alpha\left(  0\right)
+\sum_{l=1}^{\infty}\alpha\left(  i\xi_{l}\right)  I\left(  \xi_{l}\frac
{2z}{c},\frac{c}{2z~\omega_{p}}\right)  \right] \,, \label{Eq:VBabb}%
\end{equation}
where the atomic dynamic polarizability is convoluted with
$
I\left( \zeta ,\chi \right) =\left( 1+\zeta ^{2}\chi ^{2}\right) \Gamma \left( 3,\zeta \right) +\zeta ^{4}\chi \Gamma \left( 0,\zeta \right) -3\zeta ^{2}\chi \Gamma \left( 2,\zeta \right) +2\zeta ^{4}\chi ^{2}\Gamma \left( 1,\zeta \right) -\zeta ^{6}\chi ^{2}\Gamma \left( -1,\zeta \right)
$ at
Matzubara frequencies
\[
\xi_{l}={2\pi}/{\hbar} \, k_{B}T~l,~~l=0,1,2,...,
\]
 $\Gamma\left(  n,\zeta\right)  $ being the incomplete gamma function.
In addition to recovering various limiting cases, Eq.(\ref{Eq:VBabb}) also
accounts for realistic properties of conducting wall (described by plasma
frequency $\omega_{p}$.)

Atomic properties enter the atom-wall interaction through the dynamic electric-dipole
polarizability of imaginary frequency $\alpha(i \omega)$. For the two clock levels the perturbation of the clock frequency may be
expressed in terms of  the difference   $\Delta \alpha(i \omega) =
\alpha_{^3\!P_0} (i \omega) -  \alpha_{^1\!S_0} (i \omega)$.
We carried out calculations of $\alpha(i \omega)$ and $\alpha(i \omega)$ for Mg, Ca,
Sr, Yb, Zn, Cd, and Hg atoms.  We used the {\em ab initio}
relativistic configuration interaction method coupled with many-body perturbation theory.
The summation over intermediate states entering the polarizability was carried out using the Dalgarno-Lewis method.
Details of the formalism may be found in Refs.~\cite{BelDzuDer09,DerPorBab09}. Detailed dynamic polarizabilities $\alpha(i \omega)$ and $C_3$ coefficients for the ground states of alkaline-earth atoms
may be found in Ref.~\cite{DerPorBab09}.

\begin{figure}[h]
\begin{center}
\includegraphics*[scale=1.00]{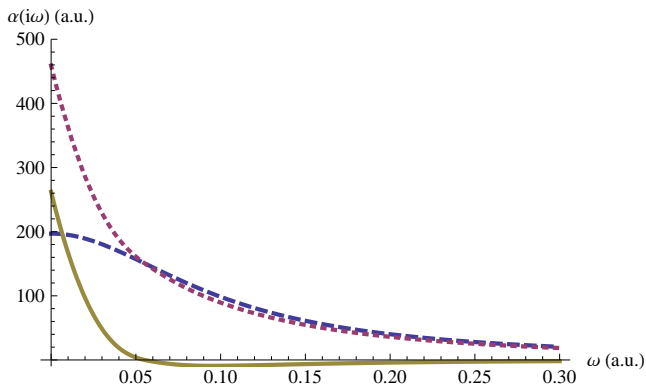}
\caption{(Color online)  Dynamic polarizabilities of imaginary frequency $\alpha(i \omega)$ of the Sr clock levels, $5s5p\,^3\!P_0$ (dotted line) and $5s^2\,^1\!S_0$ (dashed line) as a function of frequency.
Differential polarizability $\Delta \alpha(i \omega) =
\alpha_{^3\!P_0} (i \omega) -  \alpha_{^1\!S_0} (i \omega)$ is shown with a solid line. All quantities are in atomic units.
\label{Fig:difalphaSr}}
\end{center}
\end{figure}

Dynamic polarizabilities of Sr atom are shown in Fig.~\ref{Fig:difalphaSr}.
Notice that the individual polarizabilities $\alpha_{^3\!P_0} (i \omega)$ and  $\alpha_{^1\!S_0} (i \omega)$
slowly decrease as $\omega$ increases.
At large frequencies each polarizability approaches {\em the same} asymptotic limit $\alpha(i \omega) \sim N_e/\omega^2$,
$N_e$ being the number of atomic electrons.  As a result, compared to the individual $\alpha(i \omega)$,
the differential polarizability,  $\Delta \alpha(i \omega)$, is strongly peaked around $\omega=0$.
Only the Matsubara frequencies  inside this peak are relevant in Eq.~(\ref{Eq:VBabb}).
In this regard, it is worth noting that
when evaluating  $C_3$ coefficients for individual levels with Eq.~(\ref{Eq:C3}), it was found~\cite{DerJohSaf99} that neglecting
core excitations while computing $\alpha(i \omega)$ could substantially underestimate $C_3$ for heavy atoms.
Here we deal with the {\em differential} shift and a simpler approach of summing over few
first valence excitations does provide the dominant part of the effect.
Curiously, $\Delta \alpha(i \omega)$ passes through zero at
$\omega \approx 0.05 \, \mathrm{a.u.}$
This is reminiscent of the ``magic'' frequency for the perturbation of the clock transition
by  laser field which is expressed in terms of differential polarizability of {\em real} argument,
$\Delta \alpha(\omega)$.

\begin{figure}[h]
\begin{center}
\includegraphics*[scale=1.00]{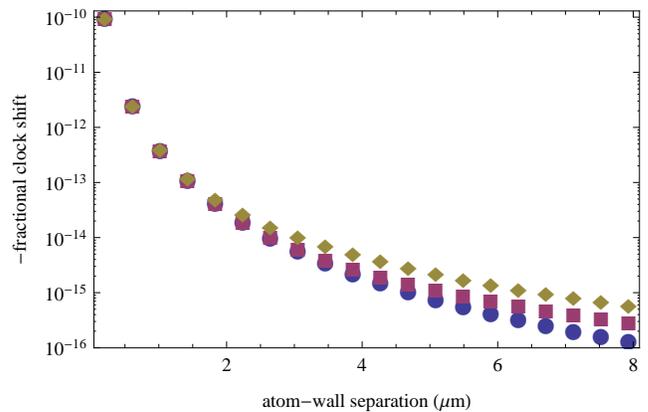}
\caption{(Color online)  Fractional clock shifts for Sr as a function of separation from a gold surface at three temperatures, $T=77\,\mathrm{K}$ (blue dots), $T=300 \,\mathrm{K}$ (red squares), and
$T=600 \,\mathrm{K}$ (brown diamonds).   Individual points represent shifts in individual trapping sites of the optical lattice. First well is placed at $\lambda_m/4$ and subsequent points are separated by $\lambda_m/2$.
\label{Fig:SrClockShift}}
\end{center}
\end{figure}

With the computed $\Delta \alpha(i \omega)$, we evaluate the atom-wall clock shifts, Eq.~(\ref{Eq:VBabb}). We use plasma
frequency $\omega_p = 9 \, \mathrm{eV}$ (gold wall) and consider several temperatures
$T= 77 \, \mathrm{K}, 300 \, \mathrm{K}$, and $600 \, \mathrm{K}$.  Results for Sr lattice clock are shown in Fig.~\ref{Fig:SrClockShift}.  Individual points represent shifts in individual wells of the optical lattice. First well is placed at $\lambda_m/4$ and subsequent wells are separated by $\lambda_m/2$.
Roughly the first 20 wells produce a fractional clock shift above the already demonstrated $10^{-16}$ accuracy limit~\cite{LudZelCam08etal}. We observe that over 20 wells the clock shift varies by six orders of magnitude. As temperature of the surface is increased, the clock shifts become more pronounced.

It is worth pointing out that
Eq.(\ref{Eq:VBabb}) assumes that the temperatures of the environment and the wall are the same (otherwise see Refs.~\cite{AntPitStr05,ObrWilAnt07}.)
Moreover, the clock shifts in Fig.~\ref{Fig:SrClockShift} do not include the conventional black-body-radiation shifts ( $\sim T^4)$. The corresponding temperature coefficients are tabulated in Ref.~\cite{PorDer06BBR}.

\begin{figure}[h]
\begin{center}
\includegraphics*[scale=1.00]{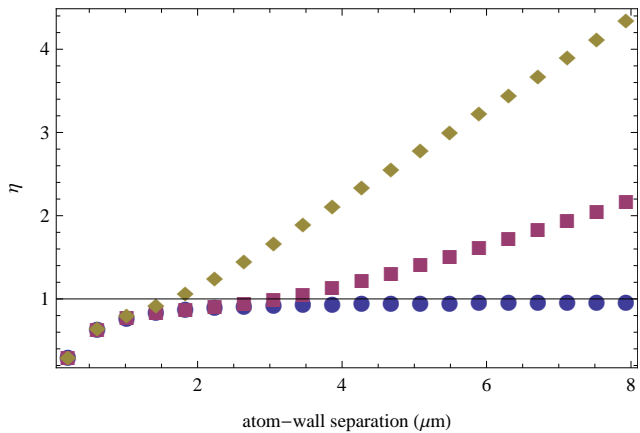}
\caption{(Color online) Sr clock shift of Fig.~\ref{Fig:SrClockShift} normalized to the Casimir-Polder limit,
Eq.~(\ref{Eq:eta}) at $T=77\,\mathrm{K}$ (blue dots), $T=300 \,\mathrm{K}$ (red squares), and
$T=600 \,\mathrm{K}$ (brown diamonds).
\label{Fig:SrEta}}
\end{center}
\end{figure}

Lattice clocks are sensitive to long-range atom-wall interactions in all three regimes: van der Walls, retardation (Casimir-Polder), and thermal-bath (Lifshitz) regimes. Indeed,
in Fig.~\ref{Fig:SrEta} we draw a ratio
\begin{equation}
\eta(z, T) = U(z,T)/U_{CP}(z) \, . \label{Eq:eta}
\end{equation}
Parameter $\eta$ is equal to one in the region where the Casimir-Polder approximation is valid.
From Fig.~\ref{Fig:SrEta}, we observe that the transition between the van der Waals
and the CP regimes occurs around well number four. The position of the second transition
region, from  the CP to the Lifshitz regimes depends on the temperature.
For $T=77$ K, this crossover is delayed until well number 25 (not shown on the Fig.~\ref{Fig:SrEta}). $T=600$ K
represents another extreme, as the van der Waals region immediately transforms  into
the Lifshitz region. Atom-wall interaction at room temperature, $T=300$ K, represents an intermediate case, where
the Casimir-Polder approximation is valid over several wells, and all the three domains
become distinguishable.

We shown that the lattice clocks can be used to detect
all three qualitative-distinct mechanisms of the atom-wall interaction.
In this regard, the lattice clocks offer a unique
opportunity to map out both van der Walls$\rightarrow$Casimir-Polder and Casimir$\rightarrow$Polder-Lifshitz transition regions. This distinguishes the clocks from
previous experiments: the former transition was probed in Ref.~\cite{SukBosCho93}, while the latter was detected in Ref.~\cite{ObrWilAnt07}.
None of the experiments so far has been able to map out both transitions simultaneously.

The accuracy of determination of the atom-wall interaction is affected by
how well the position of the clouds with respect to the
surface is determined. At each well of the optical lattice, the atomic center-of-mass
wavefunction is spread over some distance $\Delta z$. Then the clock shift acquires
a width, leading to an uncertainty
$\delta U(z,T)/ U(z,T) \approx 3 \Delta z/ z \approx
6/N_w \, \eta_{LD} \, (\lambda_{\rm clock}/\lambda_{m})$, where
$\eta_{LD}=\lambda_{\rm clock}^{-1} (\hbar/(2M\omega_{ho}))^{1/2}$
is the Lamb-Dicke parameter for an atom of mass $M$ and $\omega_{ho}$ is
the harmonic oscillator frequency of the trapping potential along the $z$-axis.
$N_w$ is the site number counting from the surface. For a typical value of $\eta_{LD}\approx 0.1$,
the error in
determination of the atom-wall interaction would be in the order of a few per cent.

The accuracy can be potentially improved  by increasing the intensity of the lattice laser, $I_L$, as the size of atomic cloud $\Delta z \propto 1/I_L^{1/4}$. At the same time,  the factors limiting the maximum of intensity  relate to the performance of the clock itself. The major factors here are hyperpolarizability (fourth order AC Stark shift) and photon scattering rate, which scale as $I_L^2$ and $I_L$, respectively. Depending on a specific trapping site, the intensity of the lattice laser could be optimized to attain a better accuracy. Finally, we notice that the present scheme could be extended to recently proposed microMagic lattice clocks~\cite{BelDerDzu08Clock}, operating in a more convenient microwave domain.

We thank J. Babb, J. Weinstein,  H. Katori, and A. Cronin  for discussions and Issa Beckun for drawing Fig.1.
This work was supported in part by the US National
Science Foundation, by the Australian Research Council and by the US
National Aeronautics and Space Administration under
Grant/Cooperative Agreement No. NNX07AT65A issued by the Nevada NASA
EPSCoR program.


\end{document}